\documentstyle[aps,preprint,epsf]{revtex}
\newcommand{\dt}{\Delta t}
\newcommand{\dw}{\Delta \omega}
\newcommand{\Tr}{\mbox{\rm Tr}}

\newcommand{\beq}{\begin{equation}}
\newcommand{\eeq}{\end{equation}}
\newcommand{\beqa}{\begin{eqnarray}}
\newcommand{\eeqa}{\end{eqnarray}}
\newcommand{\bra}[1]{\langle #1 |}
\newcommand{\ket}[1]{| #1 \rangle}
\newcommand{\kket}[1]{ #1 \rangle}

\newcommand{\MeV}{\mbox{\rm\, MeV}}
\newcommand{\GeV}{\mbox{\rm\, GeV}}

\newcommand{\fm}{\mbox{\rm\, fm}}
\newcommand{\ignore}[1]{}
\newcommand{\qu}{\mbox{$\underline{q}$}}

\begin{document}
\title{A Monte-Carlo Approach to Zero Energy Quantum Scattering\thanks{
  This work was supported in parts by the DFG Graduiertenkolleg
  'Starke Wechselwirkung' Erlangen-Regensburg, by the Bundesministerium
   f\"ur Forschung und Technologie (BMFT) and the Department of Energy
   Grant DE-FG02-91ER40608 (S.L.).}
}
\author{Stefan Lenz}

\address{
  Center for Theoretical Physics, Sloane Physics Laboratory,\\
  Yale University, New Haven, CT 06520-8120, USA }

\author{Hubertus Mall }
\address{
 Institut f\"ur theoretische Physik 3, Universit\"at Erlangen,\\
 Staudtstr. 7, 91058 Erlangen, Germany}
\maketitle
\begin{abstract}
   Monte-Carlo  methods for zero energy quantum scattering are
   developed. Starting from  path  integral  representations
   for scattering  observables, we present results  of numerical
   calculations for potential scattering and
    scattering off a schematic
   $^4 \rm He $ nucleus.
   The convergence properties of
   Monte-Carlo algorithms for scattering systems
   are analyzed using stochastic differential
   equation as a path sampling method.
\end{abstract}
\newpage
\section{Introduction}
Quantum scattering problems play an important role in many
branches of physics. Almost all our knowledge about
composite microscopic systems is obtained from
scattering experiments. The theoretical analysis of
these experiment is particularly difficult if the
target has internal degrees of freedom which couple strongly
to the projectile.
Examples of  this situation are antiproton-nucleus and
$K^-$-nucleus scattering.
Since nucleon-antiproton
or nucleon-$K^- $ scattering length is of the same order
of magnitude as the typical internucleon distance \cite{Dover,Eis,Hueff},
one expects that nuclear scattering of these projectiles
is dominated by multiple scattering correlations involving
many nucleons.

The standard approach to low energy nuclear multiple scattering is the
construction of effective optical one-body potentials \cite{Foldy}.
It has been
possible to describe succesfully the phenomenology of
elastic scattering and inelastic reactions to specific channels
\cite{Dover,Eis}
as well as exotic atoms \cite{Batty} in the framework of optical potentials.

One essential shortcoming of the optical model is that it is very
restricted from the point of view of nuclear dynamics. Phenomenological
difficulties in the description of the strong annihilation
in antiproton-nucleus  and
$K^- $ nucleus scattering are attributed to missing dynamical
properties of simple optical potentials \cite{Friedman}.
Recently it has been shown that a model which is dynamically
richer than the optical potential is
able to produce the observed large annihilation \cite{FLDS,Slenz2}.

This example shows that the development of alternative methods for
multiple scattering problems is necessary to understand important
aspects of nuclear reactions. For many-body problems with a discrete
spectrum considerable progress has been achieved by using
Monte-Carlo techniques \cite{Negele,Kalos}. They have been
applied to
quantum field theories \cite{Creutz}, atomic nuclei \cite{Carl,Lang},
quantum liquids \cite{Kalos1,Whitlock,Kalos2,Helium} and other problems
in condensed matter physics \cite{Gubernatis,Linden}.
However, for scattering problems with
a many-body system as target, standard Monte-Carlo methods are not applicable.
The fact that scattering wave functions are not normalizable, and that
the spectrum is continuous, requires the development of methods
which are specifically designed for scattering problems.

In this paper, methods are presented, which allow for the
calculation of scattering observables at vanishing projectile
energy. Their convergence
behaviour is estimated using stochastic
differential equations as a path sampling method.
A schematic many-body problem is discussed in order to demonstrate
that the methods developed for potential scattering are
applicable to more complicated systems.

This paper is organized as follows: In section \ref{StartIt}
three different path integral representations of scattering
observables at zero energy are derived. In section \ref{PathIt}
numerical algorithms are developed from these path
integral representations and applied to scattering off a
Gaussian potential. The convergence properties for scattering
problems are compared with systems which posses a bound state.
In section \ref{HeIt}
scattering off a schematic $^4\rm He$ nucleus
is treated. A discussion of the results is given in
\ref{EndIt}.

\section{Scattering observables and the path integral}
\label{StartIt}

\subsection{Path integrals}
\label{Pathes}

In this section
we briefly review
the path integral formalism following the presentation in \cite{Feynman}.
The Monte-Carlo methods developed in this paper are based on a
path integral representation of the imaginary time propagator:
\[
   \rho (\vec x_f , \vec  x_i | \beta) =
   \bra{\vec x_f } \exp( - \beta H ) \ket{\vec x_i }
\]
For a Hamilton operator $H$ which is the sum of kinetic energy and
a local potential $V(x)$
\[
   H=H_0+V(\vec  x)  = \frac{\vec p^2}{2m} + V(\vec x)
\]
one finds from
the Trotter product formula \cite{Feynman}
\beqa
    \rho(\vec x_f , \vec x_i | \beta )
       & = &  \lim_{N\to\infty}
    \prod_{j=1}^{N-1} \int d^3 x_j \left( \frac{m}{2 \pi \epsilon}
    \right)^{3/2} \exp (-S[x])
   =:    \int Dx \exp (-S[x])\label{eq:Sd}  \\
  S[x] & = &  S_0 [x] + S_V [x] \qquad
  S_0 [x] = \sum_{n=0}^{N-1} \frac{m}{2\epsilon}
           (\vec x_{n+1} - \vec x_n )^2 \qquad
  S_V [x] = \sum_{n=0}^{N-1} \epsilon V(\vec x_n )
    \label{eq:pfad},
    \label{Lagrange}
\eeqa
with $\vec x_0 = \vec x_i$, $\vec x_N = \vec x_f$ and
$\epsilon = \beta/N$.

Using  the  spectral  decomposition of the propagator one sees
that the ground  state wave function can be extracted from
it in  the  limit $\beta\to\infty$.
For a system with a discrete spectrum one finds:
\beq
  \rho(\vec x_f , \vec x_i | \beta )  =
  \sum_{n=0}^{\infty}
  \psi_n (\vec x_f) \psi_n^{\dagger} (\vec x_i)
  \exp( - \beta E_n)
  \stackrel{ \beta \to \infty}{ = }
  \psi_0 (\vec x_f) \psi_0^{\dagger} (\vec x_i)
  \exp( - \beta E_0)
  \label{Eigen}
\eeq
$\psi_n$ are energy eigenstates of the Hamiltion operator
$H$ and $E_n$ the corresponding eigenvalues.
Therefore solving the high dimensional integral (\ref{eq:Sd})
yields information
about the ground state of a physical system. Numerical
algorithms for calculating the integral (\ref{eq:Sd})
can be formulated using importance sampling techniques \cite{Kalos}.
The convergence rate in $\beta$ in (\ref{Eigen})
is controlled by the energy gap between the ground state and
the first excited state.
Excited states decay exponentially.
This is a typical
property of systems with a discrete spectrum.

The spectrum of a scattering system in absence of a bound state
is purely continuous, i.e. there is no energy gap. Like in the case of
a discrete spectrum the imaginary time propagator can be written
in terms of a complete set of eigenfunctions of the Hamiltonian
\cite{Gelman}
\beqa
   \rho(\vec x_f , \vec x_i | \beta ) & = &
   \int d^3 k \bra{\vec x_f } \kket{ \psi_k^+ } \bra{ \psi_k^+ }
   \kket{\vec x_i}
   \exp( - \beta \frac{\vec k^2 }{2 m} )
   \stackrel{ \beta \to \infty}{ = }
   \left( \frac{2 \pi m}{\beta} \right)^{3/2}
   \psi^{\dagger}_{k=0} (\vec x_i ) \psi_{k=0} (\vec x_f )
   \label{eq:Welle}.
\eeqa
The states
$\ket{\psi_k^{+} }$ are scattering wave functions with
projectile energy
$E=\vec k^2 / (2m) $. They are
solutions of the Lippmann-Schwinger
equation \cite{Newton}:
\[
   \ket{\psi_k^{+}} = \ket{\vec k}
   + \frac{1}{E-H_0+i\epsilon} V \ket{\psi_k^+},
\]
where $\ket{\vec k}$ is a momentum eigenstate.
Alternatively we could have used solutions $\ket{\psi_k^- }$ with ingoing
wave boundary conditions
\[
   \ket{\psi_k^{-}} = \ket{\vec k} +
   \frac{1}{E-H_0-i\epsilon} V \ket{\psi_k^-},
\]
or an arbitrary linear combination of these functions
in equation (\ref{eq:Welle}).

{}From equation (\ref{eq:Welle}) one sees that
for systems with a continuous
spectrum the ground state can also be found
from the large $\beta$ limit of the imaginary time propagator.
In numerical calculations two closely related problems occur:
The scattering wave functions are not normalizable and due
to the lack of an energy gap excited states are only suppressed
like $1/\beta$. Therefore long imaginary times
are necessary. In a Monte-Carlo algorithm this causes large fluctuations
which have to be supressed by generating many path samples.
In order to use Monte-Carlo methods for systems with a continuous
spectrum it is particularly important to construct estimators for
physical observables with
a small variance in order to avoid long computation times.
In the following sections we derive relations between the
imaginary time propagator and scattering observables
which can be used to find estimators for these observables.

\subsection{The cross section at $E=0$}
\label{Gelman}

In \cite{Gelman}, equation (\ref{Eigen}) is used
together with the asymptotic form of the wave function at $E=0$
\beq
  \psi_{k=0}^+ (r) \stackrel{r \to \infty}{ = } 1
  + \frac{a}{r} ; \qquad
   \sigma(E=0) = 4 \pi a^2,
   \label{eq:Waveass}
\eeq
to derive a variational principle for the scattering length $a$ and
the cross section $\sigma$. For a numerical calculation of the
scattering length with Monte-Carlo methods, the
following relation between
scattering length and scattering wave function \cite{Newton}
is more suitable as it does not require the limit $r\to\infty$:
\beq
   a  =
   - \frac{2m (2 \pi)^3 }{4 \pi}
     \lim_{k\to 0} \bra{\vec k} V \ket{\psi_{k=0}^+ }.
    \label{eq:fT}
\eeq
With this expression one can show that the cross section at $E=0$
can be found from a matrix element of the imaginary time propagator
\beq
 \sigma (E=0) =  \lim_{\beta \to \infty}
  (2\pi m\beta)^{1/2}2 \beta
  (2\pi)^3 \bra{k=0} V e^{-\beta H} V \ket{k=0}
  \label{eq:SdE} \label{eq:sigma}  .
\eeq
This can be expressed as a path integral
by using (\ref{eq:Sd}):
\[
  (2\pi)^3 \bra{k=0} V e^{-\beta H} V \ket{k=0} =
  \int d^3 x_f d^3 x_i
  V(x_f ) V(x_i)
  \int_{\vec x_i , 0}^{\vec x_f, 0} Dx \exp ( -S[x] )
\]
This result is generalized to scattering off a target with
internal degrees of freedom in section \ref{HeIt}.

\subsection{The phase shift}
\label{Phase}
An alternative path integral representation of the scattering length
can be found from the equation \cite{Gelman,Ma}
\beq
   \Delta Z(\beta)  = \Tr e^{-\beta H} - \Tr e^{-\beta H_0} =
   \sum_{l=0}^{\infty}
       \frac{2l+1}{\pi} \int_{0}^\infty dk
       \exp(-\beta\frac{k^2}{2m})
       \frac{\partial \delta_l (k)}{\partial k},
  \label{eq:TrPhase}
\eeq
which relates
the scattering phase shifts $\delta_l (k)$ of a system with
its statistical properties. Like (\ref{Eigen}), this equation holds
if there are no bound states of the projectile.
Note that both traces in the equation (\ref{eq:TrPhase})
are infinite for a system with a purely continuous spectrum.

Using the Trotter product
formula we find again a path integral representation for (\ref{eq:TrPhase})
\beqa
  \Delta Z (\beta) &=& \int Dx \exp( -S[x] )  \nonumber  \\
  S [x] & = & S_0 [x] - \ln ( e^{-S_V [x] } - 1 )
  \label{eq:DZ}
   .
\eeqa
The paths are periodic in $\beta$, i.e. $\vec x_N = \vec x_0$.
Problems which are related to the logarithm in the exponent of (\ref{eq:DZ})
will be discussed later.
In the limit $\beta\to\infty$ one finds from the threshold behaviour
of the scattering phase shifts
$\lim_{k\to 0}\delta_l (k) = a_l k^{2l+1}$
that $\Delta Z$ is proportional to the
s-wave scattering length $a=a_0$
\beq
 a = \lim_{\beta\to\infty} \pi \left( \frac{2\pi m}{\beta} \right)^{1/2}
     \Delta Z(\beta)
     \label{eq:DZa}
 .
\eeq

\subsection{The scattering length and the virial theorem}
\label{Virial}

An alternative method for calculating of the scattering
length can be found from the
scale transformation properties of the transition matrix $T$.
The transition matrix is defined as
\beq
   T(E) \ket{\vec k} = V \ket{\psi_k^+ }
   \label{Tdef}
   .
\eeq
One can show that \cite{Lenz2}
\beqa
   \frac{d}{dp} p \bra{\vec p_f} T(E) \ket{\vec p_i} & = &
   - \bra{\psi_{p_f}^{(-)}} o(\vec x)
     \ket{\psi_{p_i}^{(+)}} \label{eq:Scale1}\\
   p & = & | \vec p_f |  =  | \vec p_i |  = \sqrt{2 m E} \\
   o(\vec x) &=& 2V(\vec x) + \vec x \,  \cdot\vec\nabla V(\vec x)
{}.
\eeqa
A proof of this formula is given in appendix \ref{sec:scale}.

With this equation one can derive a path integral relation
for the scattering length $a$. First the spectral decomposition
of the imaginary time propagator in terms of the scattering
wave functions $\ket{\psi^+_k }$ and $\ket{\psi^-_k }$ is used:
\beq
  \frac{
  \int d^3 x \bra{\vec x_f} \exp( - \frac{\beta}{2} H )
  \ket{\vec x} o(\vec x) \bra{\vec x} \exp( - \frac{\beta}{2} H )
  \ket{\vec x_i} }{
  \bra{\vec x_f} \exp( - \beta H ) \ket{\vec x_i}
}
\stackrel{\beta\to\infty}{=}
  \left( \frac{ 16 \pi m}{\beta} \right)^{3/2}
     \bra{ \psi^{-}_{k=0} } o(\vec x)
            \ket{  \psi^{+}_{k=0} }
  \label{eq:SkDichte}.
\eeq
Because the wave function
$\ket{\psi^{+}_k} $ and $\ket{\psi^{-}_k} $
are identical at $k=0$ the  contributions from the points
$\vec x_f$  and $\vec x_i$ cancel.

With the help of the Trotter product formula the lefthand side
of the equation can be rewritten as a path integral:
\beq
  lhs. = \frac{ \int_{(\vec x_i, 0)}^{(\vec x_f, \beta)}
                Dx \exp( - S[x] ) o \left( \vec x(\frac{\beta}{2}) \right) }
              { \int_{(\vec x_i, 0)}^{(\vec x_f, \beta)}
                Dx \exp( - S[x] ) } =:
              \langle o(\vec x(\frac{\beta}{2}) \rangle
         \label{eq:SkPfad}.
\eeq
With equations  (\ref{eq:fT}) and (\ref{Tdef}) one finally
finds for the scattering  length:
\beq
       a = \frac{1}{32 \pi (2m)^{1/2}}
       \lim_{\beta\to\infty} \beta^{3/2}
       \langle o(\vec x(\frac{\beta}{2})) \rangle
      \label{eq:VV}.
\eeq

\section{Monte-Carlo evaluation of the path integral}
\label{PathIt}
\subsection{Reference potentials}
In path integral Monte-Carlo methods a path sampling algorithm
is used to generate paths which are distributed according
to a (normalized) probability weight
\beq
   P(x) = \frac{\exp( - S[x] )}{\int Dx \exp ( -S[x] )}
   \label{eq:Mass}.
\eeq
The average of an estimator $E(x)$ over the
sampled paths approaches the
expectation value
\[
    \lim_{N_s \to \infty} \frac{1}{N_s } \sum_{l=1}^{N_s} E[x_l]
    = \int Dx P[x] E[x] =: \langle E \rangle
\]
in the limit of an infinite number of samples $N_s $.
In the path integral representations in (\ref{eq:sigma}) and
(\ref{eq:DZa}), the scattering observable is not related to
an expectation value of a functional in path space but to the path
integral itself. Therefore we still have to construct estimators
for scattering observables.
Consider the functional
\beq
   E_{\tilde V}[x]  =  \exp \epsilon \sum_{n=0}^{N-1} \big(
        V(\vec{x}_n) - \tilde V(\vec x_n) \big)
        =  \exp(S_V [x])
        \exp(- S_{\tilde V} [x] )
   \label{eq:Obs}
    ,
\eeq
where
$\tilde V$
is an arbitrary function.
The expectation value of $E_{\tilde V}$ is
\beq
  \langle E_{\tilde V} \rangle  =
  \frac{ \int Dx \exp( -S [x] ) E_{\tilde V} [x] }{ \int Dx
         \exp( - S[x] )}
  =  \frac{ \int Dx \exp( - S_0 [x] - S_{\tilde V } [x] ) }{ \int Dx
         \exp( - S_0 [x] - S_V [x] )}
  .
  \label{eq:Obsexp}
\eeq
The potential $V$ in the numerator is replaced by the potential
$\tilde V$. Therefore from (\ref{eq:Welle}) one sees that in
the limit of large imaginary time
$\beta$ the expectation value of $E_{\tilde V}$ is the ratio of the
ground state wave functions of the two potentials \cite{SLenz}:
\beq
   \langle E_{\tilde V} \rangle
   \stackrel{\beta\to \infty}{=}
   \frac{ \psi_{k=0,\tilde V} (\vec x_i)
          \psi^\dagger_{k=0,\tilde V} (\vec x_f)
        }{
          \psi_{k=0,V} (\vec x_i )
          \psi^\dagger_{k=0,V} (\vec x_f)
        }
   \label{eq:Estimator1}
\eeq
If a solvable reference potential $\tilde V$ is chosen
which is similar to $V$ the variance
of $E$ will be small and the Monte-Carlo algorithm converges rapidly.

Equation (\ref{eq:Estimator1}) can be used to construct an
estimator for the ratio of the cross sections of $V$ and $\tilde V$.
If the potential $V$ has no zeros we can sample paths which
are distributed according to the exponential of a modified
action
\beq
    S^\prime [x] = S [x] - \ln V(\vec x_f ) - \ln V(\vec x_i)
 .
 \label{eq:Sprime}
\eeq
Unlike the path integral (\ref{eq:Sd}) the points $\vec x_f$ and $\vec x_i$ are
no longer fixed but also considered as integration variables.
The expectation value of the estimator
\beq
    E^\prime_{\tilde V}[x] = \frac{\tilde V(\vec x_f)}{V(\vec x_f)}
               \frac{\tilde V(\vec x_i)}{V(\vec x_i)} E_{\tilde V}[x]
\eeq
with paths sampled according to $S^\prime $ is the ratio
of the cross sections of $V$ and $\tilde V$
\beq
   \langle E^\prime_{\tilde V} \rangle
   = \frac{ \sigma_{\tilde V} }
   { \sigma_V }
{}.
\eeq
If the potential has zeros one can divide
the integral
over the endpoints $\vec x_f$ and $\vec x_i$
into parts where the potential does not chance sign
and perform a simulation for each of these parts.
In this case the ratio of cross sections is the
sum of these contributions.

An analogous estimator can be constructed for the path
integral (\ref{eq:DZ}). With the same arguments as before
one sees that the expectation value of
\beq
   E(x) = \frac{\exp( -S_{\tilde V}[x] ) - 1}
     {\exp(-S_V [x] ) -1 }
\eeq
approaches the ratio of the scattering lengths
in the limit of large $\beta$
\beq
  \langle E \rangle \stackrel{\beta\to\infty}{=}
  \frac{a_{\tilde V}}{a_V}.
  \label{eq:TrO}
\eeq

\subsection{Langevin simulation}

Recently Langevin simulation has been studied
as a formal approach to quantum mechanics and
as a method for practical calculations (see \cite{PaWu,Namiki}
for a review).
The basic idea of a Langevin simulation is to solve numerically the
stochastic differential equation
\beq
   \frac{\partial}{\partial t} \vec  x_n (t) =
   - \frac{\partial}{\partial \vec x_n}
   S[x(t)] + \vec\eta_n (t),
  \label{eq:DGL}
\eeq
where $\vec\eta_n (t)$ is a Gaussian random variable
\[
  \langle \eta_{n,i} \rangle = 0 \qquad
  \langle \eta_{n,i} (t) \eta_{m,j} (s) \rangle = 2 \delta_{nm} \delta_{ij}
  \delta(t-s).
\]
The time variable $t$ is not the physical time which
is represented by the index $n$ here but a simulation time.
The Langevin equation (\ref{eq:DGL}) describes the dynamics
of a whole path which moves under the influence of
a drift force $\vec\nabla_n S$ and
random fluctuations $\vec\eta_n(t) $.
One can show that the stochastic process $\vec x_n(t)$ is ergodic:
\beq
   \int D x P[x] E[x] = \lim_{T \to \infty} \frac{1}{T}
   \int_{0}^{T} dt  E[x(t)]
   \label{eq:Erg}
\eeq
with $P$ defined as in (\ref{eq:Mass}) and $E$ arbitrary.

The convergence properties of the Langevin equation are related to
the Fokker-Planck operator
\beq
  \hat{F} := \sum_n \frac{\partial}{\partial\vec x_n} \left(
             \frac{\partial}{\partial\vec x_n} +
             \frac{\partial S[x]}{\partial\vec x_n} \right)
  \label{eq:Fokker}.
\eeq
which is a differential operator in path space.
$P(x)$ is the stationary solution of the Fokker-Plank equation
\beq
   \frac{\partial}{\partial t} \Phi(x,t) = \hat F \Phi (x,t),
   \label{eq:FP}
\eeq
i.e. $\hat F P [x] = 0$. The Fokker-Plank equation describes the time
evolution of probability distributions of paths. If $\hat F$
has a negative discrete spectrum, then $\Phi(x,t)$ in
(\ref{eq:FP}) converges to $P[x]$ in the limit $t\to\infty$
if the initial distribution $\Phi(x,t=0)$ is not orthogonal on $P$.
In terms of the Langevin equation (\ref{eq:DGL}) this means
that the distribution function of the stochastic process $\vec x_n (t)$
in the limit $t\to\infty$ is $P[x]$, independent of the initial
distribution. The convergence rate, i.e. the typical times
$T$ needed in (\ref{eq:Erg}), is given by the first nonvanishing
eigenvalue of $\hat F$.

For a numerical application one discretizes the Langevin
time and simulates the stochastic process
\beqa
  \vec x_n(t+\dt) & = & \vec x_n(t) - \frac{\partial S[x(t)]}{\partial\vec x_n}
            \Delta t
             + \vec\dw_n(t)  \label{eq:DGLdisk} \\
  \langle \dw_{n,i}(t) \dw_{m,j} (t) \rangle & = & \left\{
  \begin{array}{lc}
        2 \Delta t \delta_{nm} \delta_{ij}, & {\rm if}\,\, t=s \\
        0, & {\rm else}
  \end{array}
  \right.
\eeqa
We start from an arbitrary initial path $\vec x_n (t=0) $
and iterate
(\ref{eq:DGLdisk}) with a sequence of random numbers
$\Delta \vec w_n (t) $.
The method (\ref{eq:DGLdisk}) corresponds to a simple
Euler scheme for the solution of ordinary differential
equations. Higher order algorithms have been studied recently
\cite{Kronfeld,Greiner,Kloeden}.

\subsection{An example: scattering off a Gaussian potential}

After having introduced the formal techniques we will now
apply these methods to scattering off a Gaussian
potential
\beq
   V(x)=V_0 \exp ( - \frac{1}{2 b^2 } \vec x^2 )
   \label{eq:Vpot}
{}.
\eeq
As the only relevant parameter of zero energy scattering is
$2mV_0 b^2 $ one can set $b=1$ and $m=1$ without loss
of generality.

For the method described in section \ref{Gelman}, the action
$S^\prime $  in equation (\ref{eq:Sprime}) is inserted
into the stochastic
differential equation (\ref{eq:DGL}). As an initial path of
we use
\[
   \vec x_n = \frac{n}{N} (\vec x_f - \vec x_i ) + \vec x_i
\]
with
$\vec x_f $ and $\vec x_i $
randomly chosen inside the potential.
Because of the additional part $\ln V(\vec x_i )+ \ln V (\vec x_f )$
in the action $S^\prime $ (\ref{eq:Sprime}),
the endpoints of the path  are subject to a linear
drift force in the Langevin simulation which moves them back inside
the potential region. Therefore they will not leave the
potential under the influence of the stochastic force $\dw $.

For numerical integration of the path integral defined
in \ref{Phase} the action (\ref{eq:DZ}) is inserted into
the Langevin equation (\ref{eq:DGL}). We obtain the stochastic
differential equation
\beq
  \frac{\partial}{\partial t} \vec x_n (t) =  -
  \frac{\partial S_0 [x(t)]}{\partial \vec x_n} -
  \frac{1}{D[x(t)]}
  \frac{\partial S_V [x(t)] }{\partial \vec x_n }
 + \vec\eta_n (t) \label{eq:DGL2} \qquad
  D[x] :=  1-e^{S_V [x] }
   .
\eeq
The logarithm in the action disappears,
as only the derivative of the action
enters into the Langevin equations.
For the potential (\ref{eq:Vpot}), which does not change sign,
$D [x] $ is always positive ($V$ repulsive) or negative
($V$ attractive) as $\exp S_V [x] $ is either
greater or less that one. For a general potential $D [x] $,
can have zeros. In this case the stochastic process
defined by (\ref{eq:DGL2}) is not ergodic.
The path space is divided
into regions of positive and
negative signs of $D [x] $. At the boundaries
of this
regions the drift force becomes infinite and prevents the stochastic
process from moving into another region.
{}From the numerical point of view this means that
several simulations with random initial conditions
have to be performed to make sure that the whole path
space is covered.
A more detailed interpretation of the physical
meaning of the term $1/D [x] $ in the stochastic differential
equation will be given when we discuss the convergence properties
of the algorithms. For the equation (\ref{eq:DGL2}) we use
$\vec x_n = 0$ as initial path.

An estimate of the typical times $\beta$ required in the numerical
calculation can be found from the first order Born approximation
for the matrix element
(\ref{eq:sigma}):
\beq
   \sigma \approx \left( 4\pi m \int d^3 x V (\vec x) \right)^2
   \left ( 1 - \frac{m}{\beta}
   \frac{\int d^3 x \vec x^2 V(\vec x)}{\int d^3 x V(\vec x)}
   \right)
   = (2\pi b^2)^3 (4\pi m V_0)^2 \left( 1- \frac{3 m b^2}{ \beta} \right)
\label{eq:Bornbeta}
\eeq
For the the choice of parameters $b=1, m=1$ the typical physical
times will be $\beta\approx 100$

Since analytical estimates of the
step size $\epsilon$ are in general
hard to obtain, we tested discretization errors numerically.
For the present work, we used $\epsilon=0.5$, which resulted
in errors less than 5 percent. Typically $25000$ warmup steps were
taken to equilibrate, followed by $250000$ steps of measurement
at $\dt = 0.04$. This corresponds to a Langevin time of
$T=10000$.
The choice of the
Langevin parameters is discussed in section \ref{sec:Conv} where
the convergence properties
of the algorithm are studied in more detail.

As a reference, we used a square well potential
\beq
    \tilde V(r) = \left\{ \begin{array}{cc}
    0           & r>\tilde R \\
    \tilde V_0  & r<\tilde R
    \end{array}
    \right.
    \label{eq:Vkast}.
\eeq
{}From equation (\ref{eq:Bornbeta}) we see that for
not too strong potentials, the cross section and the leading
term in $1/\beta$ is given by the zeroth and second moment of
the potential. If the parameters $\tilde R$ and $\tilde V_0 $
are chosen such that these moments are identical for both
potentials, we expect that the ratio of cross sections
is close to $1$ and depends weakly on $\beta$. This leads to
an initial guess of the parameters of $\tilde V$
\beq
    \tilde R = \sqrt{5} b \quad \tilde V_0 = 3
    \sqrt{\frac{\pi}{250}} V_0
    \label{eq:Vparam}
{}.
\eeq

\begin{figure}
\epsfxsize=14.0cm
\centerline{\epsffile{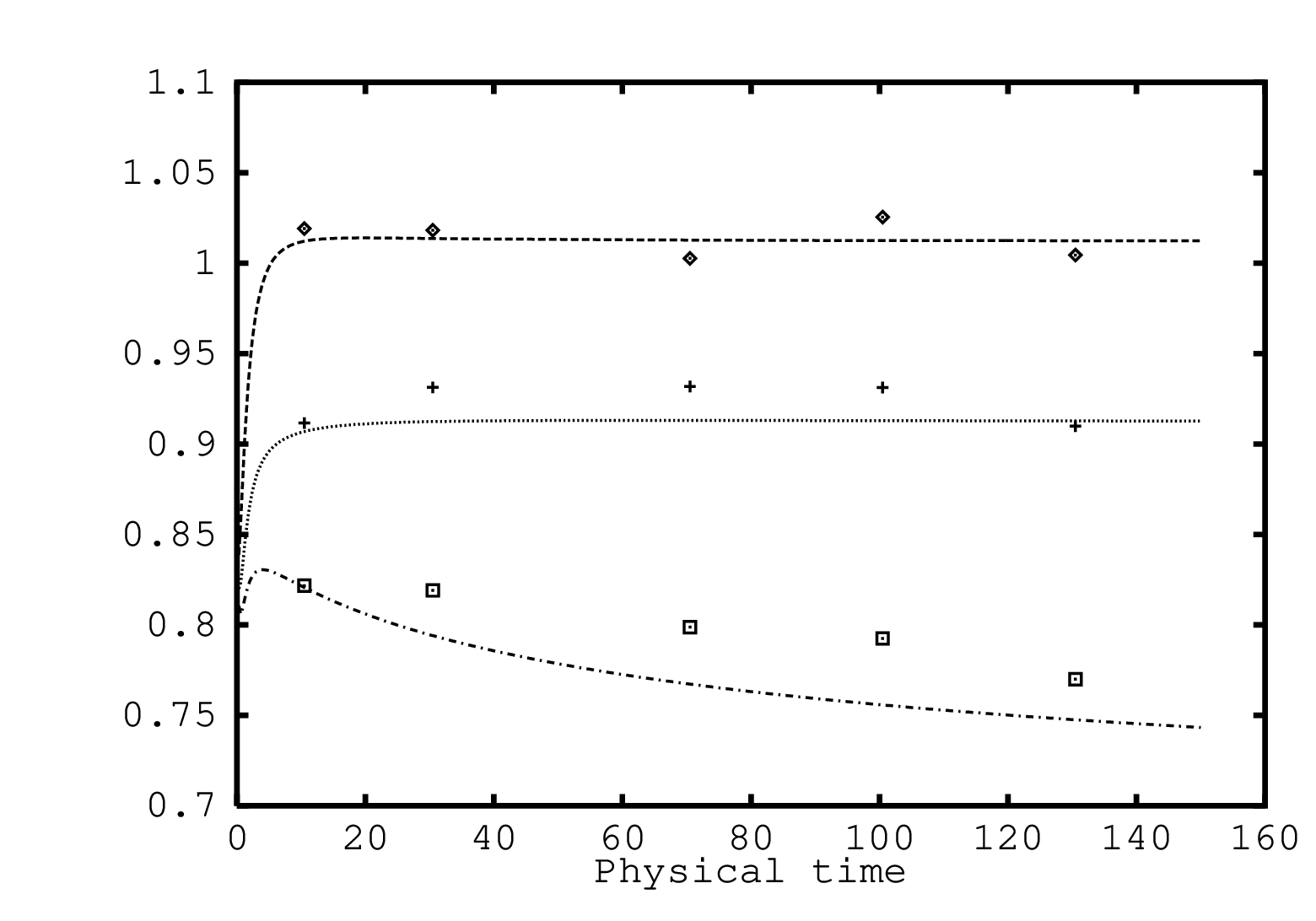}}
\caption{ Ratio of the cross sections of the
  square well potential an the Gaussian potential
  as a function of the physical time $\beta$.
  The line is the solution of the Schroedinger equation,
  data points are the results of the Langevin simulation.
  $V_0 = 0.1$ ($\Diamond$), $V_0 = -0.3$ ($+$) and
  $V_0 = -0.5$ ($\Box $) with the reference potential
  ({\protect\ref{eq:Vkast},\protect\ref{eq:Vparam}}).
}
\label{fig:Res1}
\end{figure}

We first discuss the $\beta$ dependence of the result for
the potential strengths $V_0 = 0.1$, $V_0 = -0.3$ and
$V_0 = -0.5$. Figure \ref{fig:Res1} shows
the ratio of the cross sections as a function of $\beta$. The lines
are the analytic result which are obtained by numerically solving
the Schroedinger equation for both potentials. The data point correspond
to the stochastic result.

The ratio of the cross section is predicted well
for $V_0=0.1$ and $V_0 = -0.3$ and does not depend
on $\beta$ from $\beta\approx 10$ on.
The deviation of the simulation from the exact result is
typically below 4 percent.
For $V_0 = - 0.5$ a stronger
$\beta$ dependence is found and the
simulation predicts a result which is
approximately 10 percent too large.

\begin{figure}
\epsfxsize=14.0cm
\centerline{\epsffile{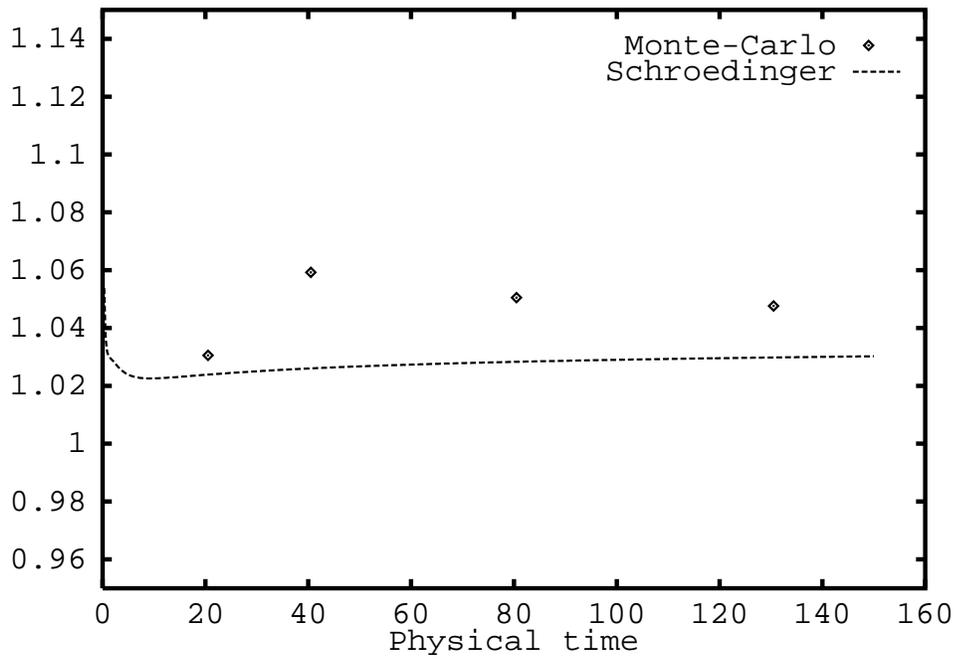}}
\caption{ The same calculation as in
 figure \protect{\ref{fig:Res1}} using a modified reference
 potential ({\protect\ref{eq:NewRef}}) for $V_0=-0.5$.
 ($\Diamond$): Monte-Carlo result, Line: Schroedinger equation.
 Note the different scale!
}
\label{fig:Res1b}
\end{figure}

Obviously the
reference potential (\ref{eq:Vkast}) which is based on
first order Born approximation (\ref{eq:Bornbeta})
is not a good model for the scattering potential if $V_0 $ is large.
The Gaussian potential has its first bound state at
$V_0\approx -0.67$, whereas the reference potential with the
parametrization (\ref{eq:Vparam}) has its
first bound state at $V_0 \approx -0.74$.
Therefore the reference potential (\ref{eq:Vkast}) is too weak
if $V_0$ is close to the bound state value.
If we choose the parameters
\beq
    \tilde R = 2 \quad \tilde V_0 = \frac{V_0}{0.67} \frac{\pi^2}{32}
    \label{eq:NewRef}
\eeq
the reference potential also has a bound state at $V_0 = -0.67$.
In figure \ref{fig:Res1b} we plot the result for this
improved reference potential at $V_0 = -0.5$, with all other parameters
unchanged. The result is now almost $\beta$
independent and deviates by not more than 4 percent from the exact
calculation (Note the different scales of fig. \ref{fig:Res1} and
fig. \ref{fig:Res1b}).
This deviation is the discretization
error in $\epsilon$.

{}From the result in figure \ref{fig:Res1b} one sees clearly
the importance of choosing a good reference potential.
With the appropriate choice of $\tilde V$ the estimator
$E_{\tilde V}$ does not strongly
depend on $\beta$ and long simulation
times can be avoided.
In practice one guesses an initial reference potential and improves
it  gradually until the variance of $E$ is minimal and no
$\beta$-dependence is observed.

\begin{figure}
\epsfxsize=14cm
\centerline{\epsffile{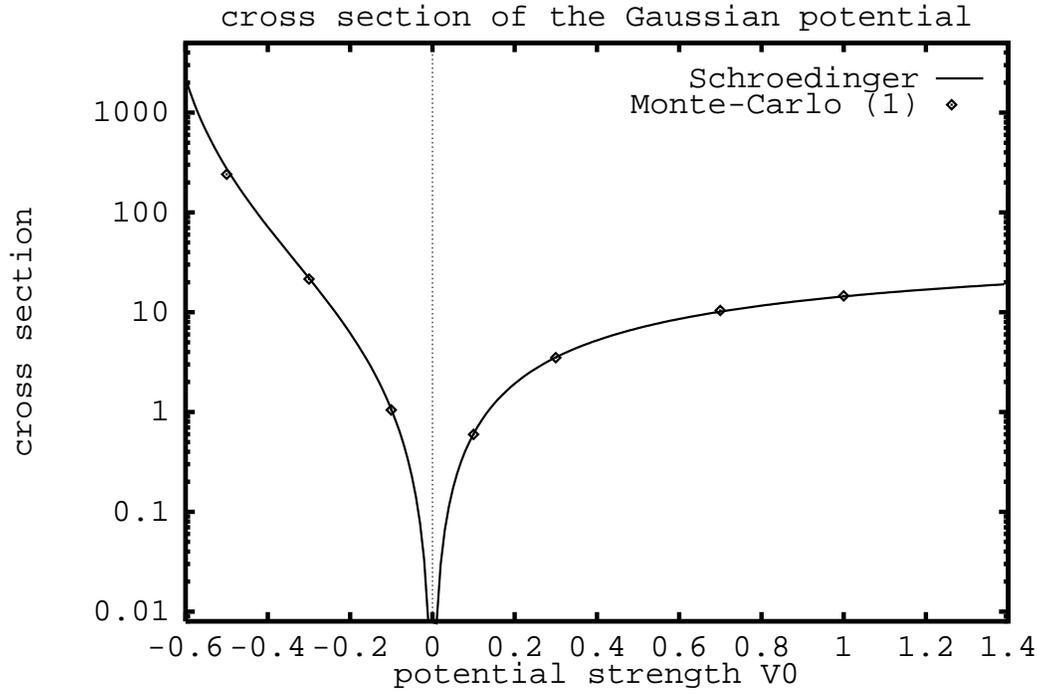}}
\caption{The cross section of the Gaussian potential
   as a function of $V_0$. The line is the result of the
   Schroedinger equation, data points are the Monte-Carlo
   simulation (\protect\ref{eq:sigma}).}
\label{fig:Res5b}
\end{figure}

Qualitatively the calculation using the method (\ref{eq:DZa})
shows the same $\beta$ dependence as in figure \ref{fig:Res1}
and \ref{fig:Res1b}.
In figure \ref{fig:Res5b} the cross section is plotted
for the simulation based on (\ref{eq:sigma}).
One sees that accurate results can be obtained
over a wide range of potential strengths and cross
sections with the same simulation parameters.

\begin{figure}[p]
\epsfxsize=14.0cm
\centerline{\epsffile{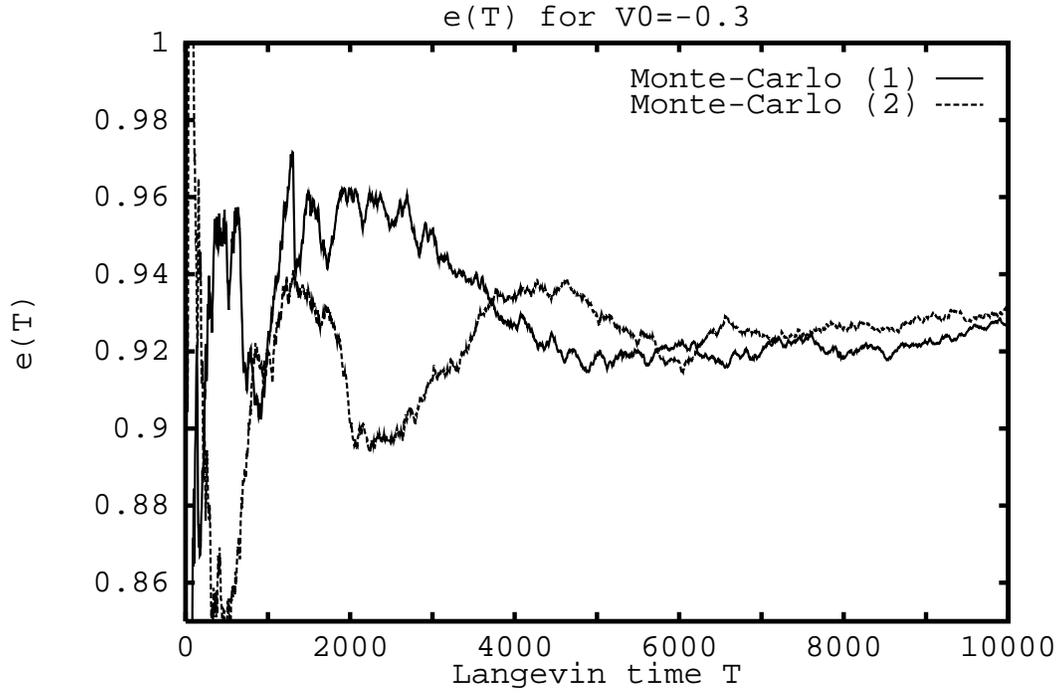}}
\caption{
   The prediction of the  ratio of the cross sections of
   the Gaussian potential
   and the reference potential as a function of the
   simulation time $T$ at $V_0=-0.3$.
   Monte-Carlo (1)  uses  ({\protect\ref{eq:sigma}}),
   Monte-Carlo (2)  uses   (\protect\ref{eq:DZa}).
}
\label{fig:ResT}
\end{figure}

The central question of every Monte-Carlo calculation
is whether the equilibrium distribution is sampled
correctly and whether
enough samples have been chosen to obtain a
statistically significant result.
The answer to this question can be found by studying the
autocorrelation time of the algorithm. For Fokker-Planck
operators with a discrete spectrum one can show
\cite{Namiki} that
for long times $t$
the correlations
of the degrees of freedom decay
like
\[
  \langle x_n (t) x_n (0) \rangle
  \stackrel{t\to\infty}{\propto} \exp( - \frac{t}{t_{ac} } )
  ,
\]
where the autocorrelation time $t_{ac} = 1/ |\lambda|$ is
the inverse of the smallest eigenvalue
$\lambda$ of the  Fokker-Planck operator.
In the next section we will use this concept to give an estimate
of the convergence properties of the Langevin simulations.
Numerically this question can be answered by studying the
dependence of the averaged estimator
\beq
      e(T) = \frac{1}{T} \int_{0}^{T} dt E^{\prime}_{\tilde V}
      [x(t)] \approx
         \frac{1}{N_s } \sum_{l=0}^{N_s } E^{\prime}_{\tilde V}[x(t_l)]
   \label{eq:OMess}
\eeq
on the simulation time $T$.
In figure (\ref{fig:ResT}) we plot $e(T)$ for $V_0=-0.3$
for both methods (\ref{eq:sigma}) and (\ref{eq:DZa}).
One sees strong fluctuations
at small $T$ which average out as $T$ increases. From $T=4000$
on the result is essentially constant.

{}From this figure one would estimate an
autocorrelation time of the order of magnitude of about $1000$,
i.e. the algorithm produces
one independent sample every $25000$ paths. This has to
be compared with typical values for bound state problems in
many body physics of about $10-100$ paths for one independent
sample (see e.g.\cite{Creutz,Gubernatis} and the references
given there). These large autocorrelation times are not a problem
of the Langevin algorithm, but a physical property of
scattering systems, as will be seen soon.

\begin{figure}
\epsfxsize=14.0cm
\centerline{\epsffile{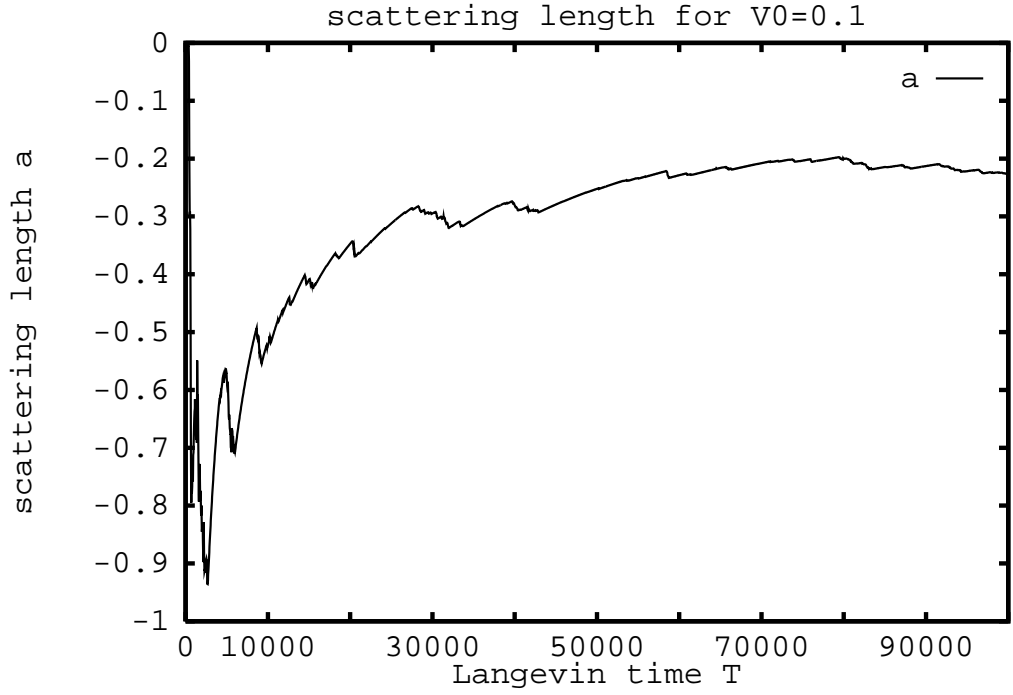}}
\caption{
  The scattering length
  $a$ for $V_0 = 0.1$ as a
  function of the simulation time $T$ at $\beta=100$.
  the exact value is $a=-0.21$.}
\label{Abb:Skala1}
\end{figure}

We now discuss the results for method (\ref{eq:VV}).
For this calculation the initial and final point of the
path are fixed to $\vec x_i = \vec x_f = 0$. The stochastic
differential equation (\ref{eq:DGL}) is used for path sampling.
In contrast to the previous methods a reference potential
is not needed as equation (\ref{eq:VV}) directly relates the
scattering length $a$ to the expectation value of the function
$o(r)$.
We again plot the result as a function of the simulation time
in figure (\ref{Abb:Skala1}).
The potential strength is $V_0 = 0.1$. Although the potential
is rather weak one needs $T=100000$, i.e. $2.5\times 10^6$
samples to come close to the analytical result of $a=-0.21$.

\begin{figure}
\epsfxsize=14.0cm
\centerline{\epsffile{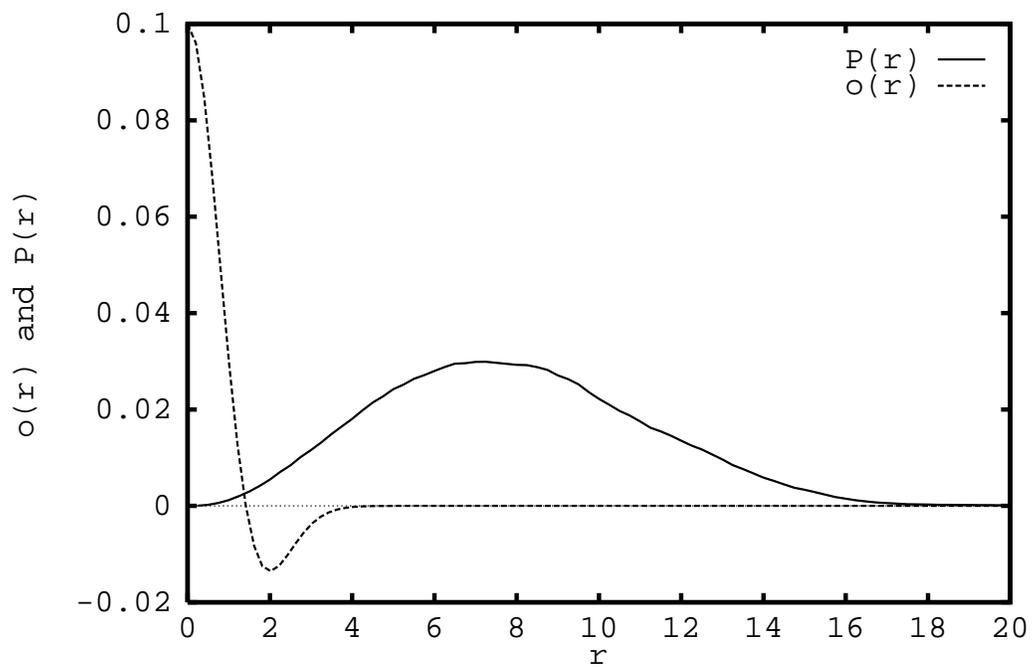}}
\caption{
 Probability distribution of path midpoints for $V_0 = 0.1$
         and $o(r)$ as a function of the distance $r$.}
\label{Abb:Skala2}
\end{figure}

The slow convergence of (\ref{eq:VV})
can be understood from the structure of the estimator $o(x)$
and the paths which are generated by the stochastic differential equation.
In figure \ref{Abb:Skala2} we plot $o(r)$ and the probability
distribution of finding the midpoint of the path at a distance
$r$ from the origin.
Obviously the midpoint of the paths lies outside
the potential most of the times, i.e. at values
of $r$ where $o(r)$ is zero. Only few samples contribute
to the expectation value $\langle o \rangle$ in (\ref{eq:VV}).
For this reason it is not advisable to calculate the scattering
length directly with eq. (\ref{eq:VV}).
It can be used together with method (\ref{eq:sigma}) to determine
the sign of the scattering length. The estimator $E^\prime_{\tilde V} [x]$ in
equation (\ref{eq:Obs}) yields accurate results for the square
of the scattering length and the same paths which are used
with $E^\prime_{\tilde V} [x]$  can be used to measure the sign of $a$
via (\ref{eq:VV}).

\subsection{Estimate of convergence properties}
\label{sec:Conv}

The numerical results presented
in the previous section showed that a rather large number
of path samples is required in the simulation. Further
it was found
that the paths are outside
the range of the potential most of the time.
Both observations are intimately related to the physical
properties of scattering systems. They are not an artifact of the
algorithm we use.

The reason for both problems is that scattering wave functions
are not normalizable.  Consider an attractive potential
which vanishes at infinity
and is strong enough to have a bound ground state at energy $E_0$.
At $\beta\to\infty$ the typical spatial extension of the paths
is $ r \approx 1/\sqrt{2mE_0}$, as the path integral is
dominated by the ground
state wave function, which decays like
$\exp(-\sqrt{2mE_0}r)$. If the potential is too weak to have a bound state
the typical extension of the paths diverges as $\beta\to\infty$,
as the zero energy scattering wave function (\ref{eq:Waveass}) does
not vanish at $r\to\infty$.
Therefore one expects that the convergence properties of the
Monte-Carlo simulation becomes worse in the limit of large $\beta$,
which means that the autocorrelation time strongly depends on $\beta$.

This
can be interpreted intuitively from the properties
of the stochastic differential equation (\ref{eq:DGL}) with the
action $S^\prime $ eq.(\ref{eq:Sprime}). For a potential
$V$, which is attractive but too weak to have a bound state,
the potential drift
$\vec\nabla_n S_V$ is not strong enough to move the paths back into
the range of the potential. If the endpoints of the paths were
not fixed inside the range of the potential by the term
$\ln V(\vec x_i ) + \ln V (\vec x_f)$ in $S^\prime$, they would move to
infinity under the influence of the stochastic force $\eta(t)$.
The other points of the paths are kept back in the vicinity
of the potential by the kinetic energy term $S_0$ of the action
which causes a linear force between neighbouring path points
in the stochastic differential equation. For systems
with at least one bound state the potential energy is strong enough
to keep the paths inside the range of the potential, even if the
endpoints are not fixed.

One can give a quantitative result for the typical behaviour of the
path points by studying only the kinetic energy part of the action plus
the part of the potential which keeps back the endpoints.
In this case the stochastic differential equation has the
form
\beq
   \frac{\partial}{\partial t}
   \vec x_n   =  - \sum_{m=0}^{N} M_{nm} \vec x_m(t)  +
                \vec \eta_n (t)
\eeq
\beqa
 { \bf  M } & = & \frac{m}{\epsilon} (
            2 {\bf 1} - { \bf A }),\quad
  {\bf A }  :=
            \left( \begin{array}{cccccc}
                    1-\frac{\epsilon}{m b^2}   & 1 & 0 & \dots         \\
                    1       & 0 & 1 & 0 & \dots     \\
                    0       & 1 & 0 & 1 & 0 & \dots \\
                    \vdots  \\
                            & \dots & 0 & 1 & 0 & 1 \\
                            &       & \dots & 0 & 1 & 1-\frac{\epsilon}{m b^2}
                   \end{array}
            \right)
  .
\eeqa
This equation describes a set of coupled harmonic oscillators.
The eigenvalues of the Fokker-Planck operator
are integer multiples of the eigenvalues of $-M$ \cite{Namiki}.
Therefore the lowest eigenvalue $\lambda$ of $M$ is related
to the autocorrelation time $t_{ac} = 1/\lambda$.

\begin{figure}[p]
\centerline{
\epsfxsize=7cm
\epsffile{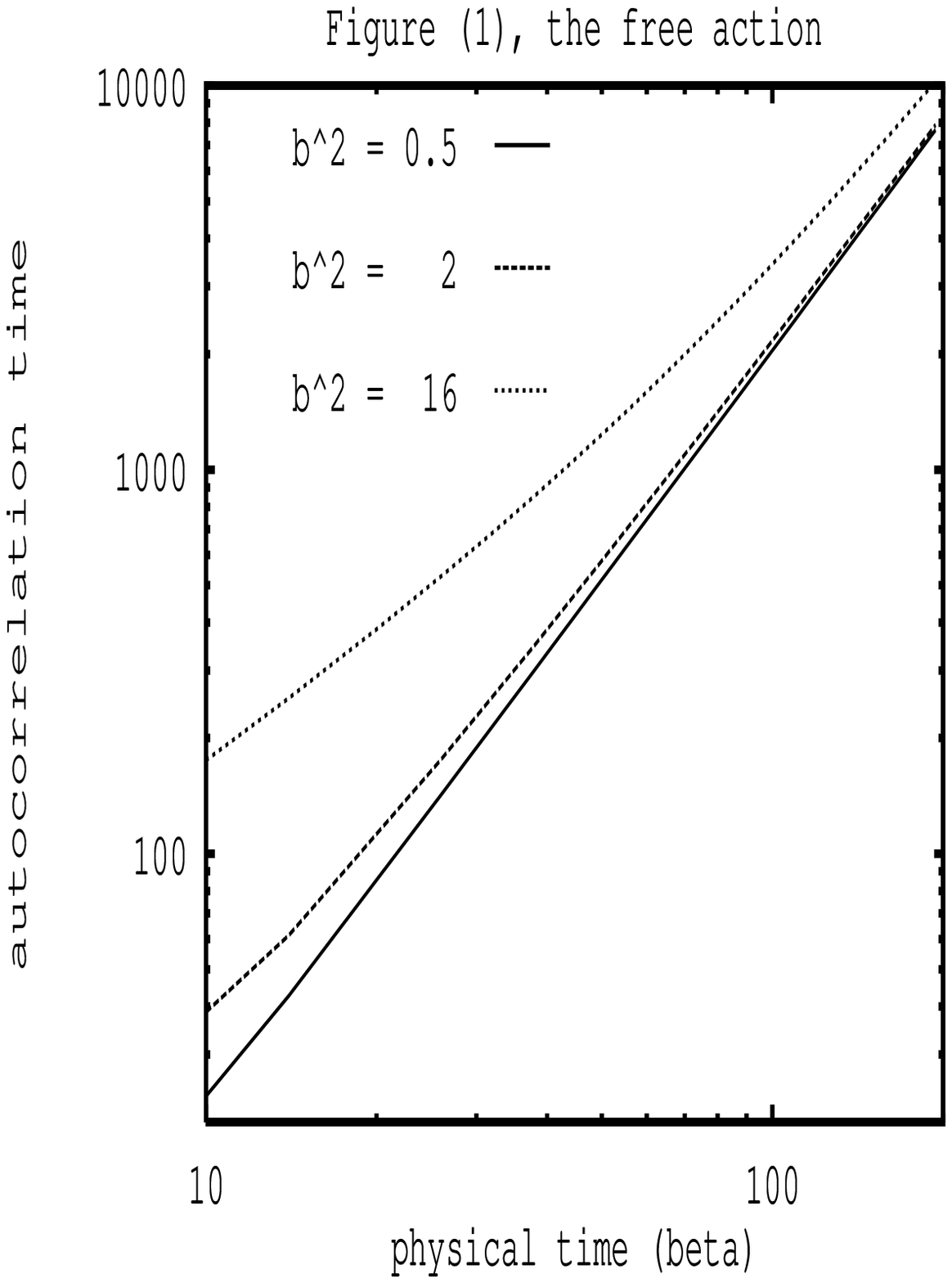}
\epsfxsize=7cm
\epsffile{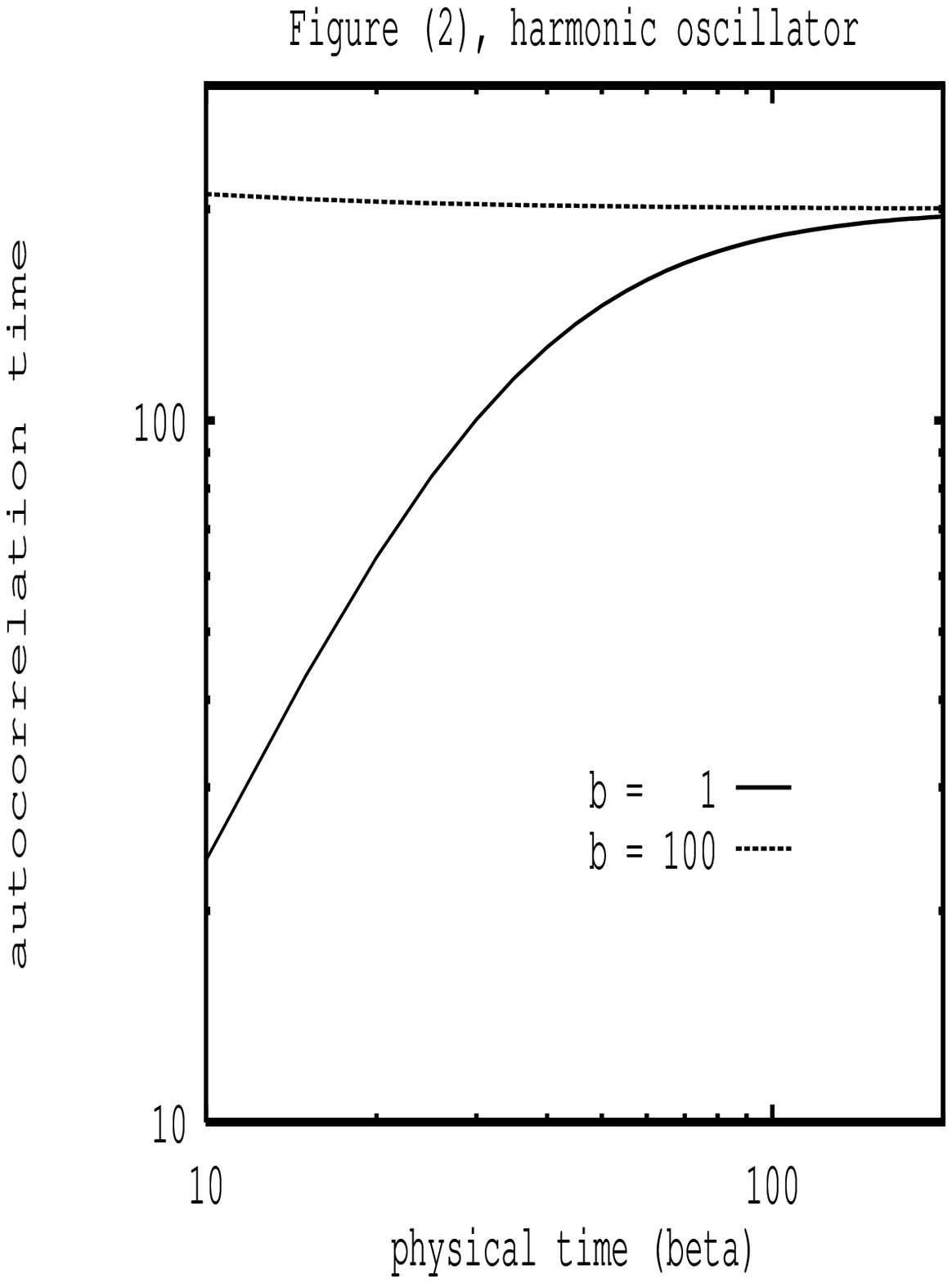}}
\caption{
   The autocorrelation time
   $t_{ac}$ for the free action.
   Figure (1): $t_{ac}$ as a function of the physical
   time $\beta$ for $\epsilon=0.5$.
   Figure (2): the free action plus a weak harmonic
   oscillator potential ($\omega=0.1$).
}
\label{fig:Acc}
\end{figure}

The eigenvalues of $M$ can be found by numerical diagonalization.
In figure \ref{fig:Acc}.1 we show the autocorrelation time
at constant $\epsilon=0.5$ as a function of $\beta$ for
different widths $b$ of the potential which binds the endpoints.
$t_{ac}$ diverges like $\beta^2 $.
For our simulation at $\beta=100$
we find an autocorrelation time of $t_{ac}\approx 2000$ which
agrees with the estimate from plotting $e(T)$
in figure \ref{fig:ResT}.

We also see from figure
\ref{fig:Acc} that the autocorrelation time depends strongly
on the width $b$.
In the limit $b\to\infty$
the endpoints of the paths are no longer fixed in the
range of the potential. In this case the matrix $M$
has one vanishing eigenvalue with the eigenvector
$ \Phi_{min} \propto ( 1 ,1 , \dots , 1)$. This zero mode corresponds
to a displacement of the path. The stochastic differential
equation is translational invariant in this case.
This causes an infinite autocorrelation time as the paths spread
over a space with an infinite volume.

Figure (\ref{fig:Acc},2)
shows the same calculation but with a small harmonic oscillator
potential
\[
   \frac{1}{2} \epsilon\sum_{n=0}^{N-1} m \omega^2 \vec x_n^2
\]
($\omega=0.1$) added to the action. Independent of the conditions
on the endpoints we reach a constant autocorrelation
time at large values of $\beta$, in agreement with our discussion
at the beginning of this section.

Therefore two conditions are necessary for the
convergence of the Langevin simulation: One is a boundary
condition on the endpoints of the paths and the other is
a finite time $\beta$. This guarantuees a discrete spectrum
of the Fokker-Planck operator. The situation is analogous
to quantum mechanics on a finite interval. The boundary conditions
on the wave function causes the spectrum to be discrete, even for
systems which have only scattering states
in the continuum limit $L\to\infty$.  The energy gap
of a system with a bound state approaches a nonzero value in this limit,
for a system with a purely continuous spectrum it vanishes.

A remarkable point in this result is that a connection exists
between the spectrum of the Fokker-Planck operator and the Hamiltonian.
Although we did not formulate a rigorous proof,
it is clear from
physical arguments that for actions of the form
(\ref{Lagrange}) the Fokker-Planck operator has a continuous sectrum
in the limit $\beta\to\infty$, if the Hamilton operator has a
purely continuous spectrum.

Let us briefly discuss the method presented in (\ref{eq:DZa})
from the point of view of boundary conditions on the paths.
We sum over periodic paths which are not fixed in space.
If a path drifts outside the range
of the potential, i.e. if $S_V$ in  (\ref{eq:DGL2})
approaches zero, the factor $1/D$ increases
the potential drift and moves the path back into the range of the
potential. This is true for repulsive and attractive potentials.
Therefore the Fokker-Planck operator of the
stochastic differential equation (\ref{eq:DGL2})
has a discrete spectrum for finite $\beta$ and the
simulation converges.

Numerically we have seen that the convergence properties
of the simulations based on (\ref{eq:sigma}) and (\ref{eq:DZa})
are very similar.
The autocorrelation time in both cases is strongly $\beta$ dependent.
The reason for this can be understood by comparing the
potential drifts of (\ref{eq:DGL2}) with the potential drift
of e.g. a particle in a harmonic oscillator potential.
For the harmonic
oscillator every point of the path experiences a linear force
moving it back to the origin, whereas in (\ref{eq:DGL2}) we
have an overall factor which only  increases the drift if most
of the points lie outside the potential.
Therefore we still have a diverging
autocorrelation time at large $\beta$.

One can also give an estimate of the maximum Langevin
discretization time $\dt$ from the eigenvalues of $M$.
This quantity is determined by the maximum eigenvalue
of $M$ as convergence of the discretized Langevin
equation requires $\lambda\dt\ll 1$ for all eigenvalues
$\lambda$. From the Perron-Frobenius theorem for
positive matrices \cite{Graham} it is found that
$4m/\epsilon$
is an upper limit for the eigenvalues of $M$.
The step size $\dt$ therefore
has to fulfill the condition
\beq
   \Delta t \ll \frac{\epsilon}{4m}
   \label{eq:lmin}
{}.
\eeq
For the parameters used here the upper bound is
$\Delta t \ll 0.125$. Numerically
strong discretization errors are found close to this value
and instabilities of the simulation above the critical value.

\section{Many-body targets}
\label{HeIt}

In the previous two sections it has been shown that for potential
scattering  it is possible
to calculate scattering observables at vanishing energy
with Monte-Carlo methods. We now give an application to
a schematic many-body system in order to show that convergence
properties are not fundamentally different for targets
with internal degrees of freedom.

Consider a target which consists of a set of $A$ harmonic oscillators
with a Hamiltonian
\beq
    H_{t} = \sum_{a=1}^{A} \left( \frac{\vec p_a^2 }{2M}
    + \frac{1}{2} M \omega^2 \vec q_a^2 \right)
  .
  \label{eq:HOHam}
\eeq
The projectile interacts with the target via a sum of local
two-body potentials
\[
   V(\vec x, \vec q_1 , \dots , \vec q_A ) =
   \sum_{a=1}^{A} V(\vec x - \vec q_a)
{}.
\]
Again the relevant object is the long time limit of the
imaginary time propagator. In the absence of a projectile
target bound state we find analogous to (\ref{eq:Welle})
\beq
   \bra{ \vec x^\prime \qu^\prime } \exp ( -\beta H )
   \ket{ \vec x \qu }
    \stackrel{ \beta \to \infty}{ = }
  \exp(-\beta E_0) \left( \frac{2\pi m}{\beta} \right)^{3/2}
   \psi_{0,0}(\vec x^\prime , \qu^\prime)
   \psi_{0,0}(\vec x , \qu )
  ,
\eeq
where $\psi_{0,0}(\vec x , \qu )$ is a scattering wave
function with the projectile at $k=0$ and the target in its
ground state $|0\rangle$ as incoming wave. $E_0$ is the
ground state energy of the target. We use the notation
$ \qu   =  (\vec q_1 , \dots , \vec q_A ) $
for the degrees of freedom of the target.

The elastic cross section of the
projectile at vanishing energy is:
\beq
   \sigma(E=0)  =  \lim_{\beta\to\infty} (2\pi m\beta)^{1/2}
   \beta e^{\beta E_0}
   (2\pi)^3 \bra{0,k=0} V e^{-\beta H} V \ket{0,k=0}
   \label{eq:CrossSec}
\eeq
The matix element can again be written as a path integral
\beqa
   (2\pi)^3 \bra{0,k=0} V e^{ -\beta H } V \ket{0,k=0}
  & = &  \nonumber \\
  \sum_{a^\prime,a=1}^{A}\int d^{3A} q_i d^{3A} q_f d^3 x_i d^3 x_f
   V(\vec x_i - \vec q_{a,i}) \Phi_0(\qu_i)
   V(\vec x_f - \vec q_{a^\prime,f}) \Phi_0(\qu_f) & \times & \nonumber \\
   \int^{\vec x_f \qu_f}_{\vec x_i \qu_i}  Dx Dq \exp(-S [x,q] ) &&
   . \label{eq:Manypath}
\eeqa
$\Phi_0(\qu)=\langle \qu | 0 \rangle$
is the target ground state wave function.

As the propagator of the
harmonic oscillator is known in closed form one can use
an improved Trotter-product formula and insert for one
time step
\beq
    \bra{ \vec x^\prime \qu^\prime } \exp ( -\epsilon H )
   \ket{ \vec x \qu } \approx
   G_0^{HO}(\qu^\prime , \qu | \epsilon )
   \left(\frac{m}{2\pi\epsilon}\right)^{3/2}
    \exp\left(  - (\frac{m}{2\epsilon}
    ({\vec x}^\prime - \vec x )^2 + \epsilon \sum_{a=1}^{A}
    V(\vec x - \vec q_a ) ) \right)
\eeq
with the imaginary time propagator
\beq
   G_0^{HO} (\qu^\prime , \qu | \epsilon )
   = \left( \frac{M \omega}{2 \pi \sinh \omega\epsilon}
     \right)^{3/2}
     \exp\left( - \frac{M \omega}{2 \sinh \omega\epsilon }
             (( \qu^{\prime 2} + \qu^2 ) \cosh \omega\epsilon
                - 2 \qu^\prime \qu ) \right)
\eeq
for the harmonic oscillator.
Using this path integral means that we have no discretization
error in the propagation of the target alone, as we summed up
the full dynamics of the target already.
Unfortunately there are, to our knowledge, no closed
expressions for propagators of scattering systems in
three dimensions.
If an exact expression for the dynamics
of the projectile could be found, only the
interplay of the projectile and target
degrees of freedom would determine the
size of the time step $\epsilon$. As the
autocorrelation time critically depends on the
number of time steps for scattering systems,
improved short time
propagators \cite{Makri,Bender} could be particularly helpful
for practical applications.

The treatment of the endpoints of the paths can be
performed in the same way as for potential scattering
by simulating the modified action
\beq
  S^\prime [x,q] =  S [x,q] - \ln \Phi_0 (\qu_i) - \ln \Phi_0 (\qu_f)
            -  \ln \sum_a V(\vec x_i - \vec q_{a,i})
                 - \ln \sum_a V(\vec x_i
		  - \vec q_{a,f} ) \label{eq:HeEnd1}
\eeq
For a many-body system the integral over the endpoints
of the paths is high dimensional and can only be handled
by stochastic integration.

In equation (\ref{eq:HeEnd1})
the logarithm of the ground state wave function is added
to the action. For the problem discussed here this is a harmonic
oscillator wave function which is analytically known. At the end
of this section we will briefly comment on how this formula can be
generalized to applications where the ground state wave function
of the target is not known.

As an estimator
a functional analogous to (\ref{eq:Obs}) can be used \cite{Slenz2}
\beq
  E[x,q] = \exp \epsilon \sum_{n=0}^{N-1} \left( \sum_{a=1}^{A}
     V(\vec x_n - \vec q_{n,a})  -  V_{\em eff} (\vec x_n) \right)
	 \label{eq:OMB}
  .
\eeq
Here the coupling of projectile and target degrees
of freedom is replaced by an effective potential $V_{\em eff}$ which only
contains the projectile degrees of freedom.
Unlike the traditional approach to nuclear
multiple scattering \cite{Eis,Hueff,Foldy} where
the many body problem is approximated by an effective optical potential,
no approximations are made here.
The Monte-Carlo method yields
an exact result and compares it with the effective
potential.

With the same arguments as for potential scattering one can show
that measuring the estimator (\ref{eq:OMB}) with paths sampled
with the modified action (\ref{eq:HeEnd1})
yields in the limit of large imaginary time $\beta$
\beq
   \langle E \rangle \stackrel{\beta\to\infty}{=} \\
  \frac{ \bra{k=0} V_g \ket{\psi_{k=0}^{V_{\em eff}} }
         \bra{\psi_{k=0}^{V_{\em eff}} } V_g \ket{k=0} }
       {
         \bra{0,k=0} V \ket{\psi_{0,0}}
         \bra{\psi_{0,0} } V \ket{0,k=0}
       }
   ,  \label{eq:expO}
\eeq
with the folded potential
\beq
  V_g (\vec x) =
  \int d^3 q |\Phi_0 (\qu)|^2 \sum_{a=1}^{A} V(\vec x - \vec q_a)
  \label{eq:Fold1}
  .
\eeq
The quantity in the denominator is the cross section of
the full many body problem whereas the numerator
can be found as the solution of a potential scattering problem.
If one uses the corrected estimator
\beq
   E^\prime  [x,q] = E[x,q] \frac{V_{\em eff} (\vec x_f )}{V_g (\vec x_f )}
                \frac{V_{\em eff} (\vec x_i )}{V_g (\vec x_i )}
   \label{eq:Oprime}
\eeq
one again finds that the expectation value of $E^\prime $
is the ratio of the cross sections
\[
  \langle E^\prime \rangle \stackrel{\beta\to\infty}{=}
  \frac{\sigma_{V_{\em eff}}}{\sigma}
{}.
\]
There
is one major difference as compared to the potential scattering
calculation. One can always find an estimator $E_{\tilde V}$ with a
vanishing variance for potential scattering by setting
$ \tilde V = V$. In other words:
we know that there are model potentials which are arbitrarily
close to the potential which we want to study. This is not true
for the estimator (\ref{eq:OMB}), because
correlations of target and projectile degrees of freedom
can not be modeled with a one-body potential.
The variance of $E$ in the simulation will show
whether this is possible or not.

As we are dealing in the following with a
system of distinguishable identical particles
we can further simplify the treatment of
the endpoint by decomposing the sum (\ref{eq:CrossSec})
into a indirect contribution $\sigma_i $, where the projectile
path begins and ends at different nucleons,
and an indirect contribution $\sigma_d$, where the projectile path begins
and ends at the same nucleon
\beqa
    \sigma(E=0) &=& A \sigma_{d} + A (A - 1) \sigma_{i} \nonumber\\
    \sigma_{d} &=& \lim_{\beta\to\infty}
     (2\pi m \beta)^{1/2}\beta e^{\beta E_0}
    \bra{0,k=0} V(\vec x - \vec q_1) e^{-\beta H}
    V(\vec x - \vec q_1) \ket{0,k=0}
     \nonumber\\
    \sigma_{i} &=&   \lim_{\beta\to\infty}
     (2\pi m \beta)^{1/2}\beta e^{\beta E_0}
    \bra{0,k=0} V(\vec x - \vec q_1) e^{-\beta H}
    V(\vec x - \vec q_2) \ket{0,k=0}.
     \nonumber
\eeqa
The advantage of this procedure as compared to (\ref{eq:HeEnd1})
is that one
has already isolated the factors $A$ and $A(A-1)$ correctly.
If one simply inserts (\ref{eq:HeEnd1}) into the stochastic
differential equation and starts the calculation with a initial
path which begins and ends at the same nucleon it might
take a long time $T$ until the simulation reached a path
which begins and ends at different nucleons.
After splitting the path integral into the direct and indirect
contribution one can perform two independent simulations
with actions
\beqa
  S_d [x,q] & = &  S [x,q] - \ln \Phi_0 (\qu) - \ln \Phi_0 (\qu^\prime )
               - \ln V(\vec x - \vec q_1) - \ln V(\vec x^\prime
		  - \vec q_1^\prime ) \nonumber \\
  S_i [x,q] & = & S [x,q] -   \ln \Phi_0 (\qu) - \ln \Phi_0 (\qu^\prime )
           -   \ln V(\vec x - \vec q_1) - \ln V(\vec x^\prime
		  - \vec q_2^\prime )  \nonumber
{}.
\eeqa
We finally get for the cross section
\beq
   \frac{\sigma}{\sigma_{V_{\em eff}}} \stackrel{\beta\to\infty}{=}
    A \frac{1}{ \langle E^\prime \rangle_d }
    +
    A(A-1)
	\frac{1}{\langle E^\prime \rangle_i}
    \label{eq:Ratio1}
{}.
\eeq
One could also specify two independent reference potentials
$V_{\em eff}^{i}$ and  $V_{\em eff}^{d}$ to define a direct and indirect
observable.

In the following this formalism is applied to scattering
of a system of four independent particles in a harmonic oscillator
potential as a simple model for a $\rm ^4 He$ nucleus.
We focus here on the convergence properties of the algorithm in
the regime of strong multiple scattering effect.
As interaction between target and  projectile
the potential (\ref{eq:Vpot}) with $b=0.6\fm$ is used.
The mass of projectile and target
are set to $m=M= 1\GeV\approx 5\fm^{-1}$.
The oscillator frequency is $\omega=0.075\fm^{-1}$ which
corresponds to a mean square radius of the nucleus of
$4\fm^2$ and an energy gap of $\Delta E = E_1 - E_0 = 15\MeV$.

For our calculation we use the folded potential (\ref{eq:Fold1})
as an effective potential
for the direct and indirect observable.
If only the ground state of the target contributes as
intermediate state the folded potential
gives the correct cross section. Therefore the deviation
of the ratio of the cross sections (\ref{eq:Ratio1}) from one
is related to the contribution of excited states of the target
in the Born series.

\begin{figure}
\epsfxsize=14cm
\centerline{\epsffile{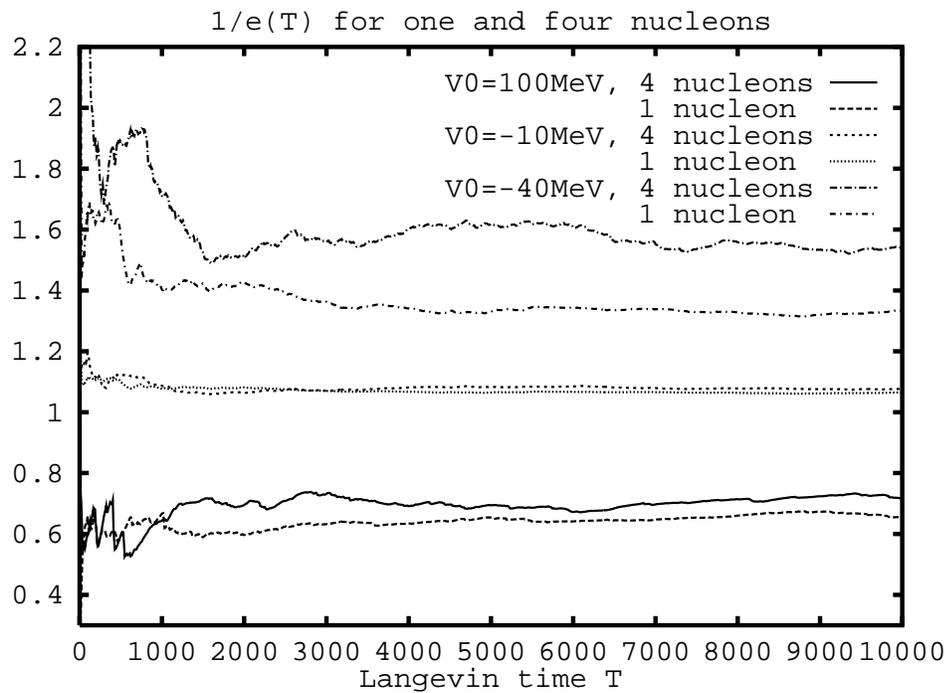}}
\caption{
  The ratio of cross sections of the full problem and the
  effective potential $1/e(T)$ as a function of the
  Langevin time $T$ for nucleon and for four nucleons.
}
\label{Testbild1}
\end{figure}

The convergence properties of the simulation can again be studied
by plotting the average of $E^\prime$ (\ref{eq:Oprime})
as an function of the simulation time $T$ analogous to eq.(\ref{eq:OMess}).
In figure \ref{Testbild1} the predicted ratio of cross sections for one
and four nucleons at $\beta=80\MeV^{-1}$ is shown.
The same discretization was chosen as for potential
scattering.

As expected the result is close to one for the weakly attractive
potential $V_0 = - 10 \MeV$ and does not fluctuate strongly. The
ground state of the target dominates the Born series and the
folded potential is a good model. The projectile-nucleus
scattering length predicted by the Monte-Carlo calculation is
$a_4 \approx 0.75\fm$ whereas the scattering length off one
bound nucleon is $a_1 \approx 0.15\fm$.
This shows that essentially only single scattering occures
as $a_4 \approx 4 a_1 $.

For the strongly attractive potential $V_0 = -40 \MeV$ the
situation is different. The projectile-nucleus scattering length
is predicted to be $a_4 \approx 240 \fm$ and the scattering length
of a single bound nucleon $a_1 \approx 1 \fm$. Therefore
for this choice of the potential multiple scattering effects
play a dominant role and the folded potential is no longer
a good model as excited states of the target become important
as intermediate states.

The statistical fluctuation in this case are larger than
for the weak potential $V_0 = -10 \MeV$. Figure \ref{Testbild1}
shows, however, that
even for $V_0 = - 40 \MeV$ the simulation has a stable
equilibrium value at $T=10000 \fm^2$.
This demonstrates that the method is well
suited for calculations where multiple scattering
effects dominate.

No qualitative differences are observed in the fluctuations
for one and four nucleons. This indicates that also simulations
with much more than four nucleons are possible, as the
computational effort scales like the number of target particles.

Finally we will briefly comment on how the treatment of the
endpoints has to be modified if the ground state wave function
of the target is not known in a closed form. In this case
one can use that the Hamiltonian of the target $H_{t}$
projects a trial state $\ket{\tilde 0}$ on the true ground state
\[
   e^{-\beta_1 H_{t}} \ket{\tilde 0}
   \stackrel{\beta_1\to\infty}{=} e^{-\beta_1 E_0} \ket{0} .
\]
We can therefore modify equation (\ref{eq:CrossSec}) by using
the expression
\beq
  \sigma(E=0)= \lim_{\beta\to\infty}
  (2\pi)^3 (2\pi m\beta)^{1/2}\beta e^{(\beta+2\beta_1) E_0}
   \bra{\tilde 0} e^{-\beta_1 H_{t}}
  \bra{k=0} V e^{-\beta H} V \ket{k=0}
  e^{-\beta_1 H_{t}} \ket{\tilde 0}
  . \label{eq:ZweiPfad}
\eeq
The time $\beta_1$ has to be chosen large on the scale
of the target dynamics. The path integral
which follows from (\ref{eq:ZweiPfad}) contains
projectile paths of length $\beta$ and target
paths of length $\beta+2\beta_1$.
In the simulation we add the logarithm of the trial wave
function to the action of the target degrees of freedom
in order to fix the endpoints. The folded potential
can also be found numerically from
\[
     V_g (\vec x) \stackrel{\beta_1 \to \infty}{=} \frac{
     \bra{\tilde 0} \exp(-\beta_1 H_{t}) V (x, \qu)
     \exp(-\beta_1 H_{t}) \ket{\tilde 0} }
     {
     \bra{\tilde 0} \exp(-2\beta_1 H_{t}) \ket{\tilde 0}
     }
 .
\]
Therefore
more complicated Hamiltonians for the target
than (\ref{eq:HOHam}) cause no fundamental
complications.

\section{Conclusion}
\label{EndIt}

The basic goal of this work was to show that Monte-Carlo
methods can be applied succesfully to scattering problems
in cases where the target has internal degrees of freedom.
One possible applications is nuclear multiple scattering, but
the methods developed here should also be useful in
atomic and molecular physics.

We showed for potential scattering and scattering of a schematic
$^4 \rm He$ nucleus that despite the
continuous spectrum of the systems one can
perform Monte-Carlo
calculations, if one uses the concept of a reference potential
to reduce statistical fluctuations.
For many-body systems a one-body reference potential
is only a first step in constructing optimal observables.
It would be useful to find solvable schematic models
which take into account the coupling of projectile and target
degrees of freedom.

For a phenomenological application of our results a number of
questions have to be answered: important candidates
for calculations in nuclear scattering are antiprotons
and $K^-$ mesons where the scattering lengths are
comparable or larger than typical internuclear
distances \cite{Dover,Eis,Hueff}.
In both systems annihilation of the projectile plays
an important role. This is modeled frequently by complex
potentials \cite{Kohno},
which are not suitable for Monte-Carlo calculations.
In order to study these problems one has to find models for
annihilation which can be implemented in a Monte-Carlo
calculation. Another open question is how to deal with the
spin of nucleons. It is well known that
antisymmetrization is very important for nucleons
as projectiles \cite{Kerman}. A related problem
is the implementation of spin-orbit interactions.
A problem of all Monte-Carlo
algorithms is, that they only converge to the ground state of
the physical system. Scattering observables can therefore
only be calculated in absence of a projectile bound state.

The main advantage
of the methods we presented is
that one can find numerical
solutions for a class of many-body problems without approximations
on the target dynamics. Questions like
in-medium effect in nuclear scattering and the reduction
to effective potentials can be investigated by purely
microscopic calculations.

A further improvement of the methods seems to be possible
in several ways: introducing improved short time propagators
(see e.g. \cite{Makri,Bender}) will increase the convergence rate of the
simulations. So far there are very few results on how
improved propagator can be constructed for scattering systems.
One of the great achievements of multiple
scattering theory is that it is formally possible to
separate elementary processes on one nucleon from the multiple
scattering dynamics \cite{Rod,Gold}. A similar formal structure
has not yet been developed in a path integral framework.

Another question is whether it is possible to design
path sampling algorithms which are optimized for scattering
problems. As scattering paths move outside the range of
the potential most of the time it might be possible to
find algorithms which use the fact that the functional
form of the solution
in the outside space is known exactly.
Langevin simulation
seems to be a promising method of developing new
algorithms because the convergence properties of the
simulation can be studied in terms of the spectrum of the
Fokker-Planck operator. For numerical applications, however,
hybrid algorithms \cite{Duane,Mac} which are based on stochastic and
deterministic dynamics in path space may be more useful.
The applicability of these methods to scattering problems
is currently studied.

\section*{Acknowledgements}

   We would like to thank Prof. F. Lenz
   who initiated this work and contributed many valuable
   ideas. Further we thank Prof. J. Koch and Prof. M. Thies
   for many discussions and T. Kraus for assistance with
   numerical algorithms for differential equations.

\appendix
\section{Proof of equation (12)}
\label{sec:scale}
Equation (\ref{eq:Scale1}) can be derived from the
scale transformation properties
of the Lippmann-Schwinger equation\cite{Lenz2}. The operator
\beq
    U = e^{ (i \vec x \vec p + \frac{3}{2} ) \ln \lambda }
\eeq
generates scale tranformations
\[
   U \vec x U^\dagger = \lambda \vec x \nonumber \qquad
   U \vec p U^\dagger =   \frac{1}{\lambda} \vec p
   .
\]
One can define a rescaled $T$-matrix and momentum eigenstates:
\beq
  \bra{\vec p_i } T(E) \ket{\vec p_f} =
  \bra{\lambda \vec p_i} T_\lambda (E) \ket{\lambda \vec p_f}
  \label{eq:99}
\eeq
\[
  T_\lambda (E)  =  U T(E) U^\dagger  \qquad
  \ket{\lambda \vec p }  =   U \ket{\vec p}
\]
As the matrix element in equation (\ref{eq:99}) does not depend
on $\lambda$ one finds:
\beq
   \frac{d}{d\lambda}
   \bra{\lambda \vec p_i} T_\lambda (E) \ket{\lambda \vec  p_f }
   = 0
  \label{eq:AnhScale1}
\eeq
{}From this equation one gets after setting $\lambda=1$
\beq
   T^\prime = \frac{1}{1-V G_0 } (2V(\vec x) + \vec x \vec \nabla
      V(\vec x) )  \frac{1}{1- G_0 V}
            - 2 T + 2 E \frac{d}{d E} T
  .
   \label{eq:AnhScale2}
\eeq
with the definition
$T^\prime = \left. \frac{d}{d \lambda} T_{\lambda} \right|_{\lambda=1}$.
For this  equation  we used that $T$  is
the solution of the operator equation
\[
   T(E) = V + V \frac{1}{E - H_0 + i\epsilon} T(E)
 .
\]
Furthermore:
\[
  2 E \frac{d}{dE} T(E) = p \frac{d}{dp} T(E(p))
  .
\]
In order to prove (\ref{eq:Scale1}) we have to write the
derivative with respect to
$\lambda$ of the momentum eigenstates as derivatives with respect to
$p$. It is easy to show that
\beqa
  \left. \frac{d}{d\lambda}
  \bra{\lambda \vec p_i} T(E) \ket{\lambda \vec p_f} \right|_{\lambda=1}
   =  p \frac{d}{dp}
    \bra{\vec p_i} T(E) \ket{\vec p_f} -
   3  \bra{\vec p_i} T_\lambda (E) \ket{\vec p_f}
  ,
  \label{eq:AnhScale3}
\eeqa
where the operator $d/dp$ only acts on the arguments of the
momentum eigenstates.
{}From the equations
(\ref{eq:AnhScale1}), (\ref{eq:AnhScale2}), (\ref{eq:AnhScale3})
and the definition of the wave functions
\[
   \ket{\psi^+_k} = \frac{1}{1-G_0V} \ket{\vec k}
   \qquad
   \bra{\psi^-_k} = \bra{\vec k} \frac{1}{1-V G_0}
\]
equation (\ref{eq:Scale1}) follows immediately.

\section{Technical remarks}

All Monte-Carlo simulation in this work were done on
an HP9000/720 workstation. A simulation at $\beta=100$
and $\epsilon=0.5$ for potential scattering with
$250000$ path samples requires approximately 40 CPU-minutes
for method (\ref{eq:sigma}).
The code is written in standard FORTRAN 77.
Method (\ref{eq:DZa}) needs about the same amount of time.
As the convergence
properties of the many body problem which we studied are
comparable to potential scattering, the computation
time scaled with the number of degrees of freedom. For the
$ ^4 \rm He$ calculation 5 hours of CPU time were needed for
one simulation run. Tests with improved random number generators
and higher order Langevin algorithms showed that an increase
of speed by a factor of 5 should be possible but we did
not study this systematically.

%
%
\newpage
\newpage
\newpage
\newpage
\newpage
\newpage
\newpage
\end{document}